\documentclass[preprint]{aastex}

\newcommand{\myemail}{cwalsh@strw.leidenuniv.nl}

\usepackage{amsmath}
\usepackage{amssymb}
\usepackage{url}
\usepackage{gensymb}
\usepackage{graphics}
\usepackage{subfigure}
\usepackage{wasysym}
\usepackage{textcomp}
\usepackage[version=3]{mhchem}

\shorttitle{DETECTION OF METHANOL IN A PROTOPLANETARY DISK}
\shortauthors{Catherine Walsh et al.}

\begin{document}

\title{FIRST DETECTION OF GAS-PHASE METHANOL IN A PROTOPLANETARY DISK}
\author{Catherine Walsh\altaffilmark{1}, 
Ryan A.~Loomis\altaffilmark{2}, 
Karin I.~\"{O}berg\altaffilmark{2}, 
Mihkel Kama\altaffilmark{1}, 
Merel L.~R.~van 't Hoff\altaffilmark{1},
Tom J. Millar\altaffilmark{3},
Yuri Aikawa\altaffilmark{4},
Eric Herbst\altaffilmark{5},
Susanna L.~Widicus Weaver\altaffilmark{6},
Hideko Nomura\altaffilmark{7}}

\altaffiltext{1}{Leiden Observatory, Leiden University, P.~O.~Box 9531, 2300~RA Leiden, The Netherlands}
\altaffiltext{2}{Harvard-Smithsonian Center for Astrophysics, 60 Garden St., Cambridge, MA 02138, USA}
\altaffiltext{3}{School of Mathematics and Physics, Queen's University Belfast, University Road, Belfast, BT7 1NN, UK}
\altaffiltext{4}{Center for Computational Sciences, University of Tsukuba, 1-1-1 Tennoudai, Tsukuba 305-8577, Japan}
\altaffiltext{5}{Departments of Chemistry and Astronomy, University of Virginia, Charlottesville, VA, 22904,USA}
\altaffiltext{6}{Department of Chemistry, Emory University, Atlanta, GA 30322, USA}
\altaffiltext{7}{Department of Earth and Planetary Science, Tokyo Institute of Technology, 2-12-1 Ookayama, Meguro-ku, 152-8551 Tokyo, Japan}

\email{\myemail}


\begin{abstract}
The first detection of gas-phase methanol in a protoplanetary disk (TW~Hya) is presented. 
In addition to being one of the largest molecules detected in disks to date, 
methanol is also the first disk organic molecule with an unambiguous ice chemistry origin. 
The stacked methanol emission, as observed with ALMA, is spectrally resolved and 
detected across six velocity channels ($>3 \sigma$), reaching a peak signal-to-noise
of $5.5\sigma$, with the kinematic pattern expected for TW~Hya.   
Using an appropriate disk model, a fractional abundance of 
$3\times 10^{-12} - 4 \times 10^{-11}$ (with respect to \ce{H2}) 
reproduces the stacked line profile and channel maps, 
with the favoured abundance dependent upon the assumed vertical location 
(midplane versus molecular layer).  
The peak emission is offset from the source position suggesting that the methanol emission 
has a ring-like morphology: the analysis here suggests it peaks at $\approx 30$~AU reaching 
a column density $\approx 3-6\times10^{12}$~cm$^{-2}$.
In the case of TW Hya, the larger (up to mm-sized) grains, residing in the inner 50~AU, 
may thus host the bulk of the disk ice reservoir.  
The successful detection of cold gas-phase methanol 
in a protoplanetary disk implies that the products of ice chemistry 
can be explored in disks, opening a window to studying complex 
organic chemistry during planetary system formation. 
\end{abstract}



\section{INTRODUCTION}
\label{introduction}

Protoplanetary disks contain the ingredients for building planetary systems. 
The disk volatile component determines the composition of the atmospheres 
of forming planets and icy planetesimals (e.g., comets).  
Whether a planetary atmosphere or icy planetesimal is e.g., carbon- or oxygen-rich,  
and/or rich in more complex molecules, will depend on radial variations 
in disk midplane composition, and specifically the locations of snowlines  
that delineate the boundary beyond which a particular volatile is frozen 
out onto dust grains \citep[see, e.g.,][]{oberg11,walsh15}.  
The Atacama Large Millimeter/Submillimeter Array (ALMA) 
can map the radial distribution of cold dust and 
gas in disks around nearby young stars with unprecedented spatial 
resolution and sensitivity.  
Recent discoveries include imaging of \ce{N2H+} and \ce{DCO+}
emission from TW~Hya, HD~163296, and IM~Lup, that show 
ring-like structures which may be tracing the CO gas distribution \citep{qi13,qi15,oberg15a}, 
the first detection of a complex organic molecule, 
\ce{CH3CN}, in MWC~480 \citep{oberg15b}, and spatially-resolved observations 
of dust gaps potentially carved by forming planets \citep{walsh14a,alma15,nomura16}.

Disk snowlines are also important for the formation of complex organic molecules 
(COMs) in disks, and thus their incorporation into nascent planets. 
Methanol, \ce{CH3OH}, is a COM of particular interest because 
it is an important parent species of larger and more complex molecules in the gas and ice  
\citep[e.g.,][]{charnley92,bennett07,oberg09a,debarros11,chen13,chuang16}.
COMs are commonly detected in the earlier, protostellar stage of star 
formation \citep[][]{herbst09}.
These molecules bridge simple species 
with those considered of prebiotic importance, e.g., amino acids.  
However, COMs in disks have proven elusive, partly because disks are small objects 
(spanning, at most, a few arcseconds)
and partly because the outer regions of protoplanetary disks are 
generally cold, dense, and well shielded from stellar radiation.  
The volatility of many COMs is similar to water ice 
\citep[e.g.,][]{oberg09a,burke14}; 
hence, COMs are expected to be frozen out onto dust grains throughout most of the disk.    
To date, only a single COM has been detected toward a protoplanetary disk: 
methyl cyanide, \ce{CH3CN}, in the disk around the Herbig~Ae star, MWC~480 ($T_\mathrm{eff} \approx 8000$~K). 
The exact chemical origin of this molecule in disks is unclear; 
modeling suggests that the emission comes from thermal or non-thermal 
ice sublimation in the comet-forming zone \citep{oberg15b}, 
which around a Herbig~Ae disk extends out to at least 100~AU. 

The most common beacon of a rich organic chemistry at the early 
stages of star formation is \ce{CH3OH}.  
Prior to this study, \ce{CH3OH} had not yet been detected in  
protoplanetary disks.  
Other O-bearing complex organic species have also remained elusive.  
By contrast, disk chemistry models that include an active 
ice chemistry as well as non-thermal desorption pathways, including photodesorption, 
predict that disks should contain large reservoirs of \ce{CH3OH} and other 
oxygen-containing COMs. 
\citet{walsh14b} determined that gas-phase \ce{CH3OH} was a good ALMA target for the indirect 
detection of the complex organic ice reservoir, reaching a peak abundance $\sim~10^{-9}$ 
(relative to \ce{H2}) in disk layers that are exposed to UV photons.  
These model calculations motivated a successful ALMA  proposal which has resulted 
in the first detection of gas-phase methanol in a nearby protoplanetary disk 
(TW Hya, $54$~pc), the results for which are reported here.  

\section{OBSERVATIONS}
\label{observations}

The young star, TW~Hya, was observed on 2015 January 02 with 39 antennas 
and baselines from 15 to 350~m (project 2013.1.00902.S, P.~I. C. Walsh). 
The quasars J1256-0547 and J1037-2934 were used for bandpass and phase calibration, 
respectively, and Titan was used for amplitude calibration.  
The Band~7 (B7) methanol transitions listed in 
Table~\ref{table1} were targeted using a channel width of 122~kHz 
(corresponding to 0.12~km~s$^{-1}$).   
A continuum-only spectral window at 317~GHz 
(with a total bandwidth of 2~GHz) was also covered. 
The total on-source observation time was 43~minutes.  
All data were self calibrated prior to imaging 
using CASA version 4.3 and the continuum 
band at 317~GHz and using a timescale of 20~seconds 
($\approx$~3 times the integration step).  
This increased the dynamic range of the continuum data by a factor of $\approx 30$.  
Self calibration using the continuum around the methanol lines between  
$\approx 304$ and 307~GHz gave almost indistinguishable results.  
The rms noise level achieved for the 317~GHz continuum following imaging with CLEAN
(Briggs weighting, robust = 0.5) was 0.10~mJy~beam$^{-1}$ 
with a peak signal-to-noise (S/N) of 4200.  
The continuum synthesized beam was 1\farcs2$\times$0\farcs6 ($-86\degree$). 
The line data were imaged without CLEANing (using natural weighting) 
following self calibration and continuum subtraction with a slight overgridding in 
velocity resolution (0.15~km~s$^{-1}$).  
The rms achieved in the dirty channel maps was $\approx4$~mJy~beam$^{-1}$ 
per velocity channel at all four frequencies.  

TW~Hya was also observed on 2014 July 19 
with 31 antennas and baselines from 30 to 650~m (project 2013.1.00114.S, P. I. K.~\"{O}berg). 
The quasar J1037-2934 was used for both bandpass and phase calibration, 
and Pallas was used for flux calibration.  
The Band~6 (B6) methanol transitions listed in 
Table~\ref{table1} were targeted using a channel width of 122~kHz 
(corresponding to 0.15~km~s$^{-1}$), 
with a total bandwidth of 58.6~MHz per spectral window.   
The total on-source observation time was 41~minutes.  
The continuum was strongly detected (peak S/N $\approx 500$), 
enabling self calibration of the B6 data with CASA version 4.3 
using the continuum within each respective spectral window and a timescale of 30~seconds.  
This increased the dynamic range of the data in each spectral window by a 
factor of $\approx 3$.
The synthesized beam was 0\farcs5$\times$0\farcs5 ($-89\degree$). 
Following self calibration, the line data were continuum subtracted 
and imaged to the same velocity resolution as the B7 data.  
An rms noise $\approx 5$~mJy~beam$^{-1}$ per channel was achieved in the dirty channel maps 
at all three frequencies.  

A significant signal was not found in the individual channel maps of the B7 data, 
nor the B6 data.   
The three B7 datasets at $\approx 304$, 305, and 307~GHz (corresponding to the lowest energy transitions) 
were concatenated into a single measurement set 
following velocity regridding and then imaged using CLEAN with natural weighting (i.e., ``stacked'' in the uv domain). 
This resulted in a synthesised beam of 1\farcs4$\times$0\farcs73 ($-84\degree$) and 
an rms noise of 2.0~mJy/beam, which is a factor of $\approx 2$ increase in sensitivity. 
A single round only of additional CLEANing was performed.  
Significant methanol emission is detected across six channels 
($\gtrsim 3\sigma$) from 2.45 to 3.35~kms$^{-1}$ reaching a peak 
S/N of $5.5\sigma$. 
No significant signal was found in the stacked and imaged B6 data.  

\begin{deluxetable}{cccc}
\tablecaption{Methanol transitions\label{table1}}
\tablewidth{0pt}
\tablehead{\colhead{Transition}&&\colhead{Frequency}&\colhead{Upper level energy}\\
\colhead{} && \colhead{(GHz)} & \colhead{(K)} }
\startdata
\multicolumn{4}{c}{Band 6} \\
\hline
$5_{0}-4_{0}$   & (E) & 241.700 & 47.9 \\
$5_{-1}-4_{-1}$ & (E) & 241.767 & 40.4 \\
$5_{05}-4_{04}$ & (A) & 241.791 & 34.8 \\
\hline \\
\multicolumn{4}{c}{Band 7} \\
\hline
$2_{11}-2_{02}$ & (A) & 304.208 & 21.6 \\
$3_{12}-3_{03}$ & (A) & 305.473 & 28.6 \\
$4_{13}-4_{04}$ & (A) & 307.166 & 38.0 \\
$8_{17}-8_{08}$ & (A) & 318.319 & 98.8   
\enddata
\end{deluxetable}

\section{RESULTS}
\label{results}

\subsection{Detection of gas-phase methanol}

\begin{figure*}[!h]
\includegraphics[width=\textwidth]{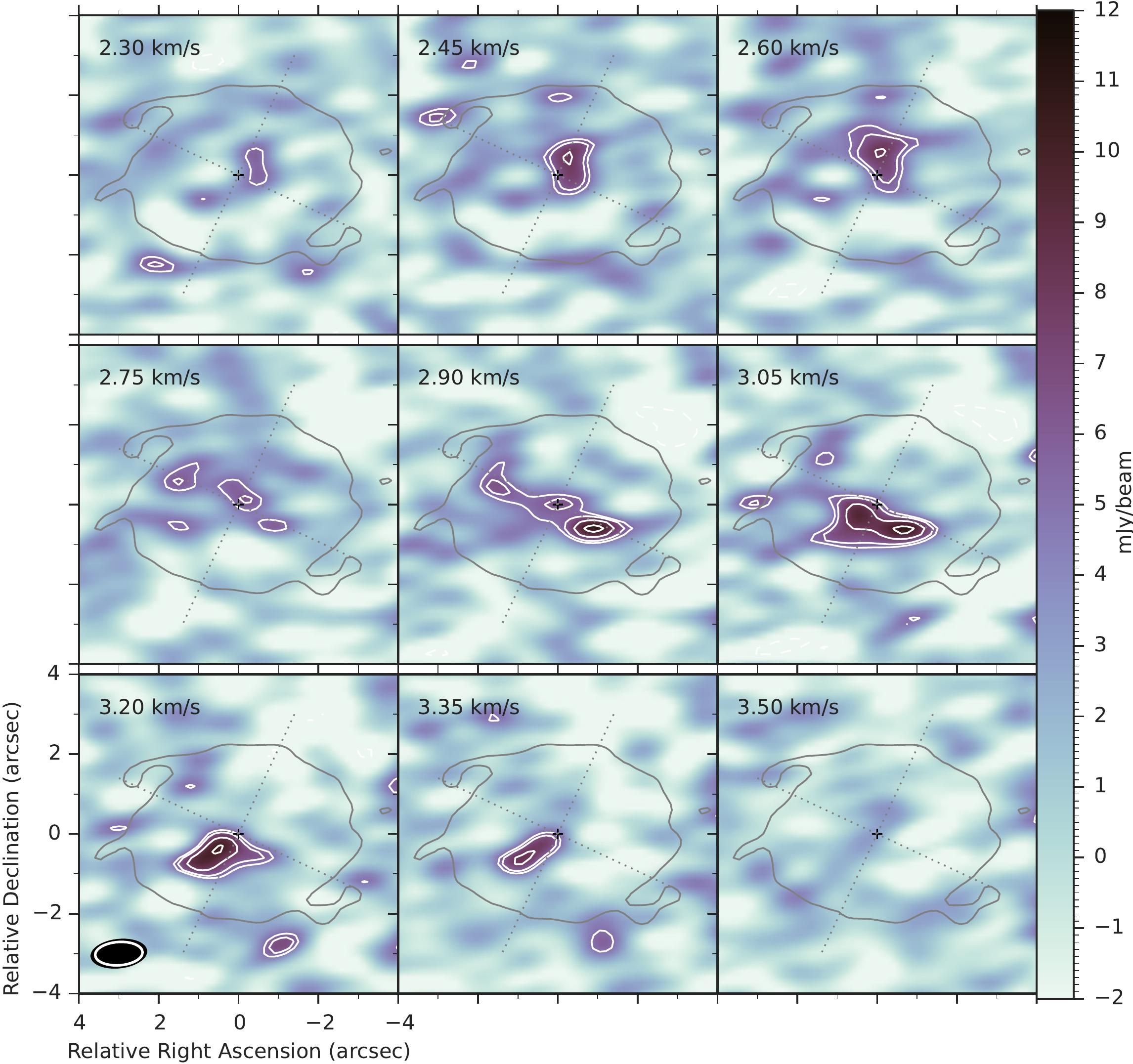}
\caption{Channel maps for the stacked observed B7 \ce{CH3OH} line emission.  
The white contours show the 2.5, 3.0, 4.0 and $5.0\sigma$~levels 
for the \ce{CH3OH} data and the gray 
contour shows the $3\sigma$ extent of the 317~GHz continuum. 
The black cross denotes the stellar position, and the dashed gray lines show the 
disk major and minor axes. 
The synthesised beams for the continuum (open ellipse) and line (filled ellipse) 
emission are shown in the bottom-left panel.}
\label{figure1}
\end{figure*}

The channel map for the stacked B7 data is presented in Figure~\ref{figure1}.  
The inclination and position angle of TW~Hya 
\citep[$7\degree$ and $335\degree$, respectively,][]{hughes11} 
are such that emission in the north-west and south-east, 
is respectively blue-shifted and red-shifted in velocity with respect to the systemic velocity 
\citep[2.9 kms$^{-1}$,][]{hughes11,andrews12}.  
This kinematic structure is evident in the methanol channel map.
In all channels in which emission is detected, the peak is offset from the 
stellar position indicating that the emitting methanol is possibly located in a ring;
however, there is significant emission ($\gtrsim 3\sigma$) at the source position.  
The detected emission appears compact compared with the extent of the continuum emission 
($\approx100\pm20$~AU). 
This may be due to the low S/N, but we cannot exclude a real drop 
in the \ce{CH3OH} abundance in the outer disk.

\begin{figure}[!h]
\includegraphics[width=0.5\textwidth]{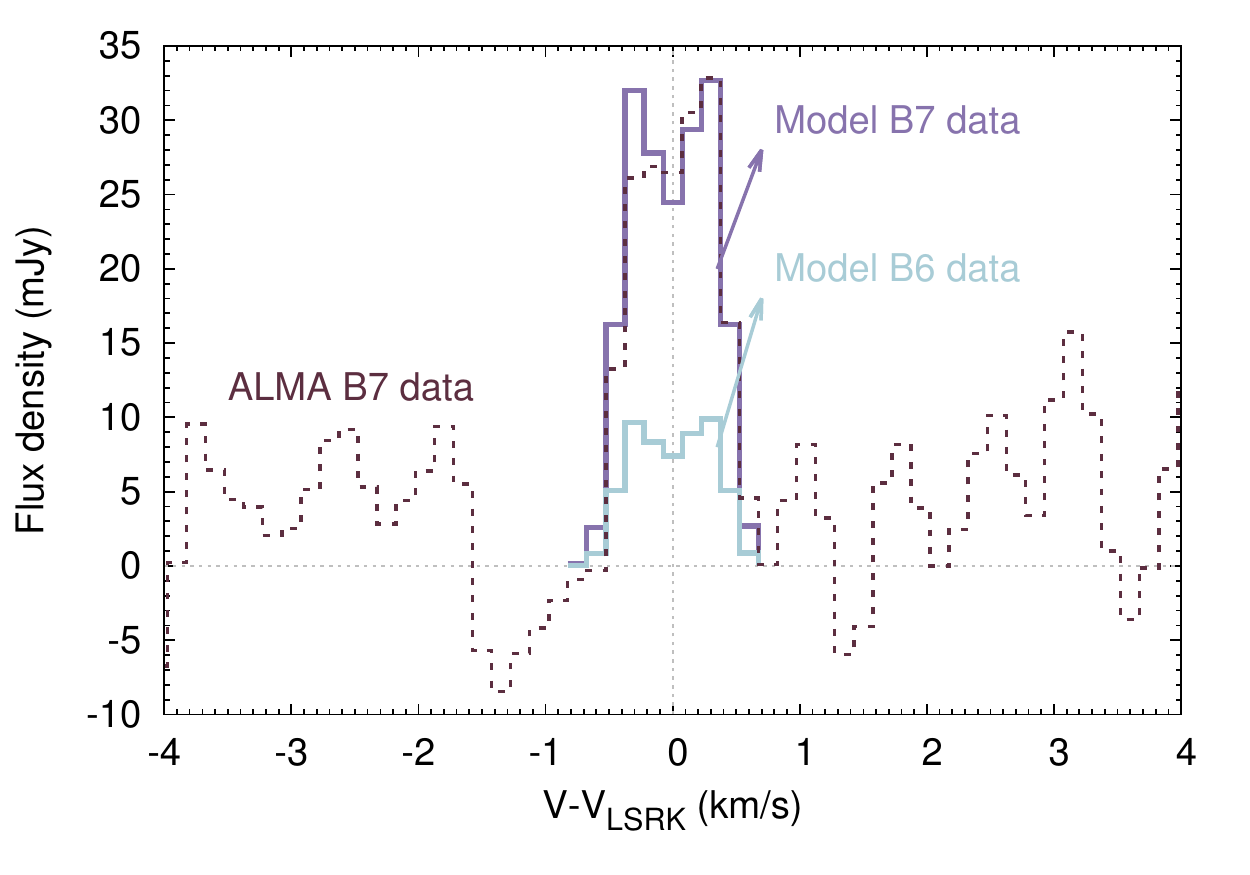}
\includegraphics[width=0.5\textwidth]{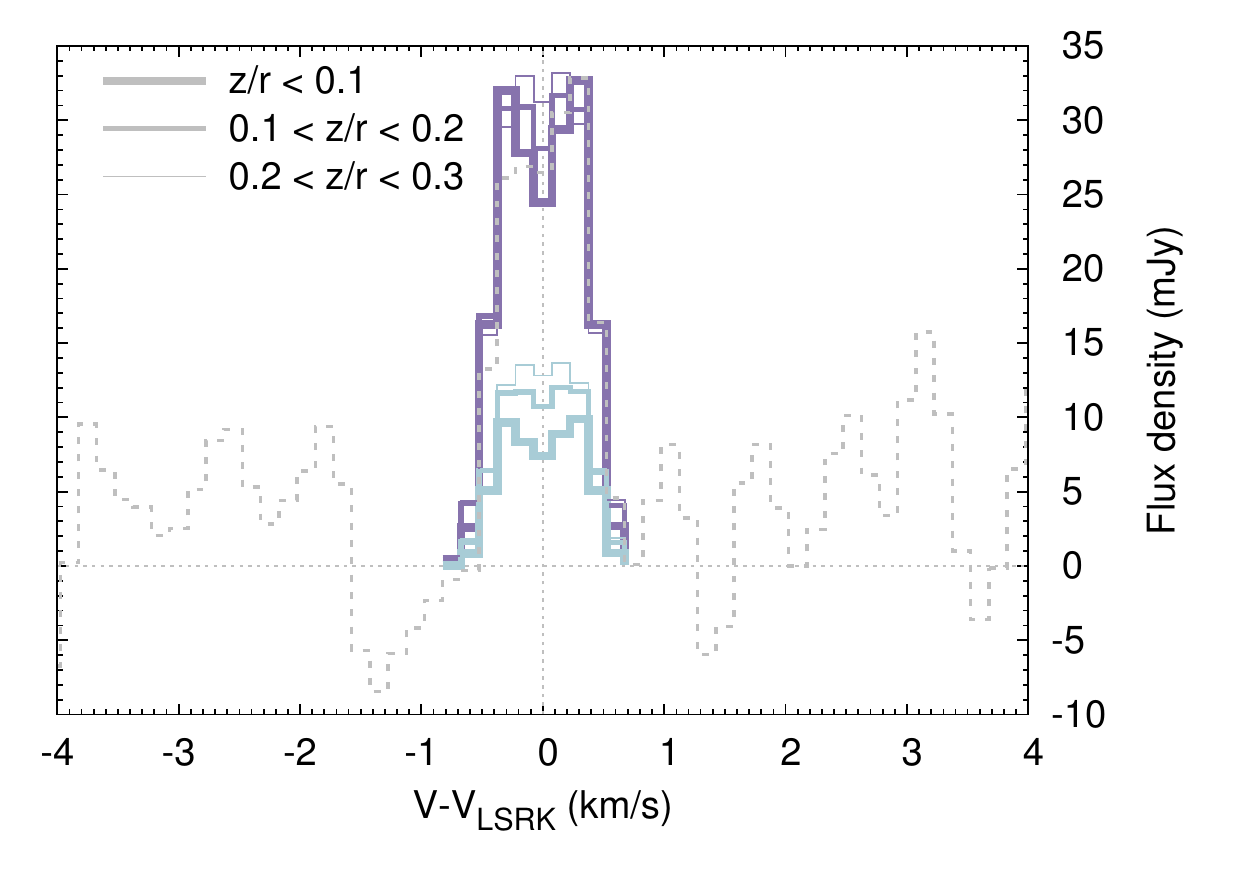}
\caption{Left: line profile extracted from the stacked observed B7 
\ce{CH3OH} channel map within the 
$3\sigma$ contour of the 317~GHz continuum (dark-red dashed lines) 
compared with the stacked B6 and B7 model line profiles 
from the best ``by-eye'' model fit for $z/r<0.1$ (light blue and purple lines, respectively). 
Right: best ``by-eye'' fits for models in which methanol is present in different layers: 
$z/r\le 0.1$, $0.1\le z/r\le 0.2$, and $0.2\le z/r\le 0.3$. 
In all models, methanol is located between $30-100$~AU.}
\label{figure2}
\end{figure}

Figure~\ref{figure2} shows the extracted line profile 
(dark-red dashed lines in left-hand panel) 
within the $3\sigma$~contour of the continuum which is marked by the grey 
contour in Figure~\ref{figure1}.  
The peak flux density for the stacked line profile is 33~mJy and 
the rms noise is 6.5~mJy, resulting in a S/N of 5.1.  

\subsection{Constraining the methanol abundance}

\begin{figure*}
\includegraphics[width=\textwidth]{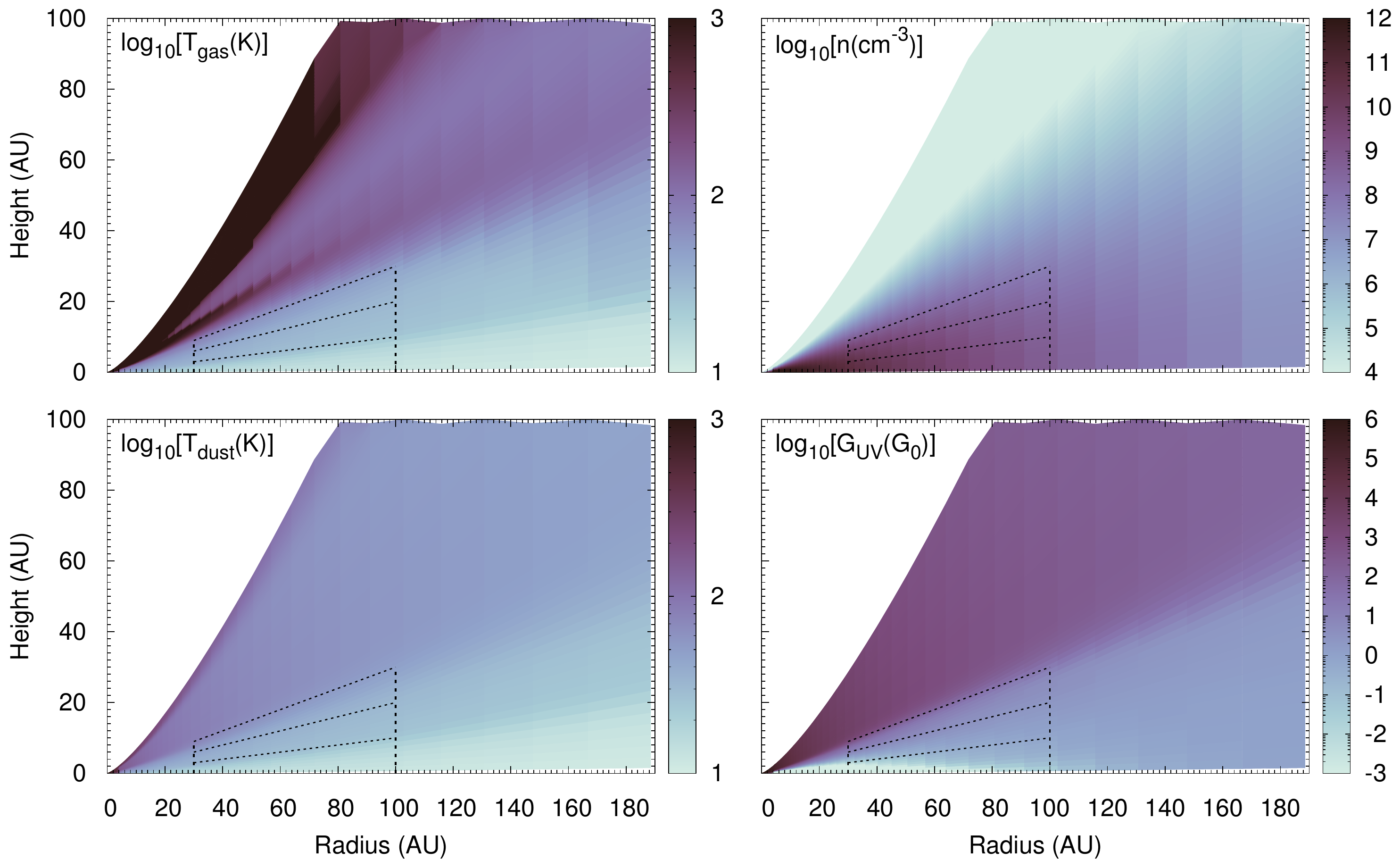}
\caption{The TW Hya disk physical structure \citep{kama16}. 
Top-left and moving clockwise: 
gas temperature (K), gas number density (cm$^{-3}$), 
UV flux (in units of the interstellar radiation field), 
and dust temperature (K).  
The dashed black line delineates the layers within which 
gas-phase \ce{CH3OH} is modeled to reside.}  
\label{figure3}
\end{figure*}

The spectrally and spatially resolved \ce{CH3OH} emission is used together with an 
appropriate density and temperature structure to constrain the 
\ce{CH3OH} abundance in the TW Hya disk.  
The low S/N of the B7 detection precludes more detailed modeling. 
The non-detection in complementary B6 data  (see Table~\ref{table1}),  
also provides additional constraints.   
We adopt the TW~Hya disk physical structure from \citet{kama16} that 
reproduces the dust SED as well CO rotational line emission from both 
single-dish observations and spatially-resolved ALMA data.  
In Figure~\ref{figure3} we show the gas temperature, 
number density, dust temperature, and far-UV integrated flux.  

Interstellar methanol is formed on or within icy mantles on dust grains 
via CO hydrogenation \citep{watanabe02,fuchs09,boogert15}.  
\ce{CH3OH} has a similar volatility to water ice \citep[e.g.,][]{brown07}; 
hence, methanol should reside on grains throughout most of the disk ($T\lesssim 100$~K).  
A small fraction of methanol can be released at low temperatures via 
\emph{non-thermal desorption} which is triggered by energetic 
photons or particles or by energy released during exothermic chemical reactions 
\citep[reactive desorption, e.g.,][]{garrod06}.  
The rates for such processes remain relatively unconstrained except for a small 
set of molecules and reaction systems 
\citep[e.g.,][]{westley95,oberg09b,fayolle11,bertin13,fillion14,cruzdiaz16,minissale16}.  
It also remains a possibility that gas-phase chemistry 
contributes to the gas-phase abundance \citep[e.g.,][]{charnley92,garrod06}. 
For example, the rate coefficient for the OH~+~\ce{CH3OH} gas-phase reaction 
was only recently determined to be rapid at low temperatures 
\citep[$<100$~K,][]{shannon13,acharyya15}.  

Because the methanol chemistry applicable to disks remains relatively unconstrained, 
we adopt a parametric approach to constraining the location and abundance 
of gas-phase methanol in TW Hya. 
Methanol is assumed to reside in three different vertical 
layers: $z/r \le 0.1$, $0.1 \le z/r \le 0.2$, and $0.2 \le z/r \le 0.3$, 
where $z$ and $r$ are the disk height and radius, respectively  
(dashed lines in Figure~\ref{figure3}).  
A small grid of models were run in which the inner and outer radius of the emission 
were varied (in steps of 10~AU), in addition to the fractional abundance of methanol 
(relative to \ce{H2}).
Ray-tracing calculations were performed using LIME \citep[LIne Modelling Engine,][]{brinch10} 
assuming LTE, a disk inclination and position angle appropriate 
for TW Hya \citep{hughes11}, and the molecular data files for A- and E-type methanol 
from the Leiden Atomic and Molecular Database 
(LAMDA, \url{http://home.strw.leidenuniv.nl/~moldata/}).  
The modeled channel maps for each individual transition 
were produced using the same spatial and spectral resolution 
as the observations and then stacked.  

\begin{figure*}[!h]
\includegraphics[width=\textwidth]{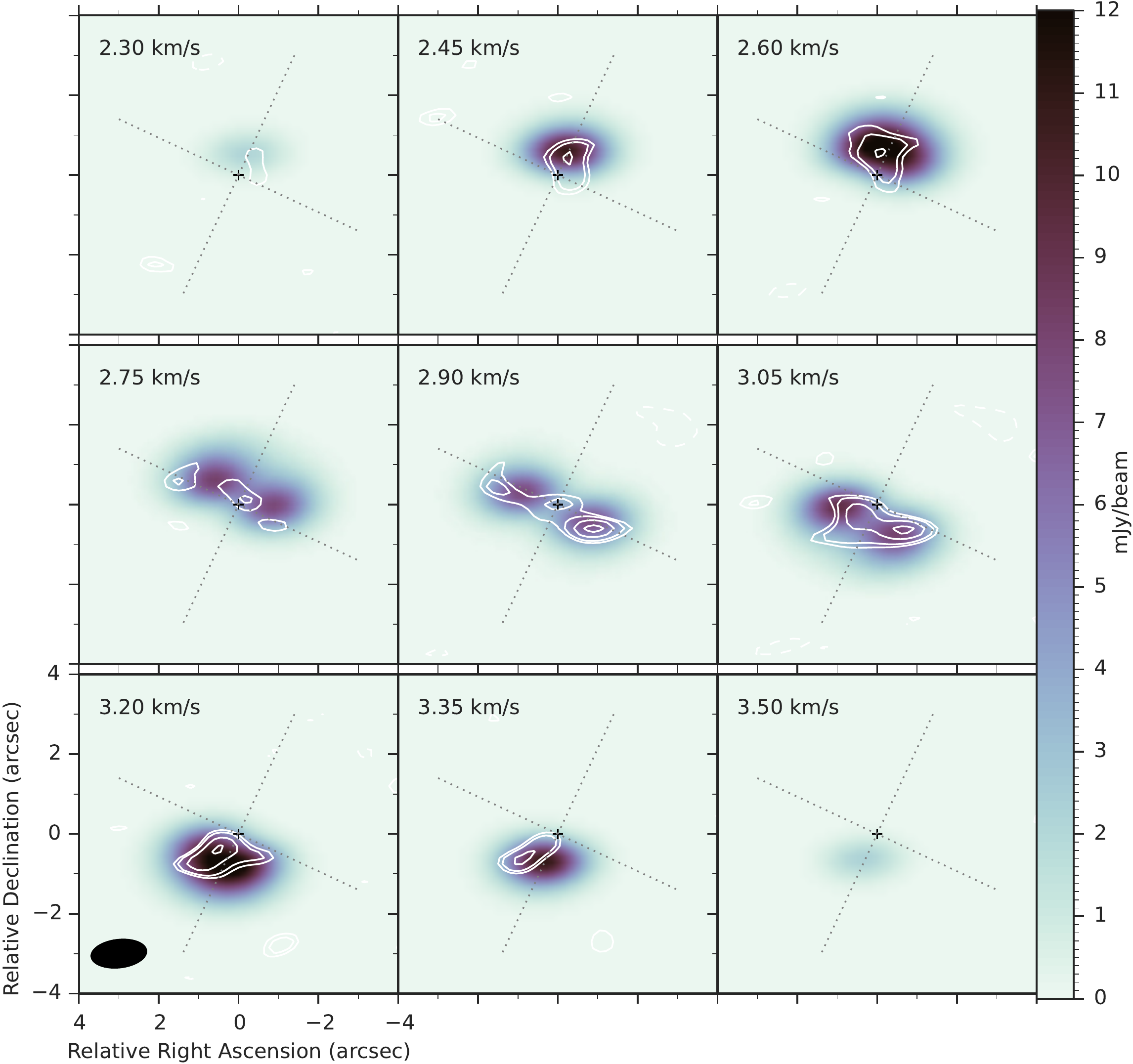}
\caption{Channel maps for the best ``by-eye'' model B7 \ce{CH3OH} line emission for $z/r\le 0.1$.  
The images have been convolved to the same beam size as the observations.
The white contours show the 2.5, 3.0, 4.0 and $5.0\sigma$~levels for the observed B7 data. 
The black cross denotes the stellar position, and the dashed gray lines show the 
disk major and minor axes.} 
\label{figure4}
\end{figure*}

The results for the best ``by-eye'' fit to the data for $z/r \le 0.1$ 
(i.e., the midplane) 
are shown in Figures~\ref{figure2} and \ref{figure4}.  
This is considered the fiducial model because the physical model 
adopted has 99\% of the dust mass in large (up to mm-sized) 
grains which are also well settled to the midplane \citep[see][for details]{kama16}.
A ring of methanol between 30 and 100~AU, 
with a fractional abundance of $2.8\times10^{-12}$ relative to \ce{H2}, 
reproduces both the shape and peak of the observed line profile, and 
the peak and radial extent of the channel maps.  
The estimated error on the abundance is $\approx 20$\%.  
If the fractional abundance of methanol is not constant,  
then the distribution may be more compact or extended than 
suggested here.

Keeping the radial location fixed and moving the methanol 
to higher layers in the disk, results in 
fractional abundances of $8.0\times10^{-12}$ and $4.4\times10^{-11}$, 
for $0.1 \le z/r \le 0.2$ and $0.2 \le z/r \le 0.3$, respectively 
(see the right-hand panel of Figure~\ref{figure2}).  
The models suggest a peak column density of 
$\approx 3 - 6 \times 10^{12}$~cm$^{-2}$ at 30~AU and 
reproduce the non-detection in the B6 data. 
Better data are required to constrain 
the vertical location of the methanol. 

\section{DISCUSSION}
\label{discussion}

Based on the presented observations, methanol is unambiguously detected in the disk of TW Hya.  
However, the deduced abundances are up to two orders of magnitudes lower than expected 
from recent disk chemistry models \citep[e.g.,][]{walsh14b,furuya14}. 
The fractional abundance with respect to \ce{H2} ranges from 
$\approx 3 \times 10^{-12}$ to $\approx 4 \times 10^{-11}$, depending 
on the location of the methanol (towards the midplane or higher 
in the molecular layer).  
Combining these values with a revisitation of the gas-phase water distribution in 
TW~Hya, suggests that the \ce{CH3OH}/\ce{H2O} gas-phase ratio is 
$\approx 0.7$\% if both species are located in the midplane, and 
$\approx 0.9-5.0$\%  if both are located higher in the molecular layer 
\citep[][]{salinas16}.  
The midplane ratio is low relative to the median interstellar 
and protostellar ice ratio \citep[$\approx 5$\%,][]{boogert15}; 
however, all values fall within the observed range for both protostars and 
comets \citep{mumma11,boogert15}.

The data and analysis also suggest that methanol is located in a ring, 
with a peak column density coincident with the location of the 
CO snow line as inferred from previous ALMA data \citep[$\approx 30$~AU,][]{qi13}.  
The emission is also relatively compact 
and peaks within the radial extent of the sub-mm dust disk.  
This supports the hypothesis that the ice reservoir in protoplanetary 
disks is hosted primarily on the larger dust grains which have 
become decoupled from the gas, and drifted radially inwards towards 
the star \citep[][]{andrews12,hogerheijde16}. 
Note that the simple modeling presented here suggests that the methanol 
emission may extend beyond the sharp drop in continuum intensity at $\approx 47$~AU 
recently discovered by \citet{hogerheijde16}.  
Higher spatial resolution data are needed to further constrain 
the radial distribution of methanol. 

The \ce{CH3OH}/\ce{H2O} gas-phase ratio relative to that expected in the ice 
can shed light on 
whether or not both species share a common gas-phase chemical origin in disks.
Possible candidates for releasing non-volatile ice species into 
the gas-phase include photodesorption by UV photons, or
reactive desorption during formation.  
Irradiation by higher energy photons and particles ($\gtrsim$~keV)  
likely induces chemistry within the ice mantle, 
rather than the desorption of intact methanol \citep[e.g.,][]{bennett07,debarros11,chen13}. 
Moreover, it has recently been shown that UV irradiation of 
pure and mixed methanol ice does not lead to the desorption of intact methanol, 
with photofragments only observed \citep{cruzdiaz16,bertin16}. 
The derived upper limit for methanol photodesorption 
($\lesssim 10^{-6} - 10^{-5}$ molecules photon$^{-1}$) is $2-3$ orders of 
magnitude less efficient than that for water ice \citep[][]{westley95,oberg09b}.  
Detailed chemical modelling will be needed to determine whether 
such a low photodesorption rate is sufficient to maintain a detectable 
level in the TW~Hya disk and reproduce the observed \ce{CH3OH}/\ce{H2O} gas-phase ratio.  
Adopting the analytical formula for the methanol abundance presented in 
\citet{furuya14}, the experimentally constrained photodesorption rate 
could account for the estimated gas-phase abundance. 
However, this should be tested using a physical model more applicable for TW Hya, 
which is colder and more settled than the T-Tauri disk structure used in \citet{furuya14}.     
Also to be explored, is whether the release of photofragments into 
the gas phase may seed methanol formation via gas-phase chemistry. 

Reactive desorption is another possibility \citep[e.g.,][]{garrod06}.  
However, given that the disk inherits the bulk of its methanol ice from 
the dark cloud and/or protostellar envelope \citep{drozdovskaya14}, in order 
for reactive desorption to occur, the methanol ice needs to be converted back 
to \ce{H2CO}, or \ce{CO}, and subsequently rehydrogenated.  
This can be achieved with UV photons 
(which likely has a low efficiency) or via surface hydrogen-abstraction reactions 
(e.g., $\ce{CH3OH}+\ce{H}\longrightarrow\ce{CH2OH}+\ce{H2}$, \citealt{chuang16}, 
and see also \citealt{martin-domenech16}).  
However, if the ice reservoir follows the mm-sized grains, 
as suggested by this study and others \citep[e.g.,][]{salinas16}, 
then the methanol ice may be concentrated towards the disk midplane, 
where both the H/\ce{H2} gas-phase ratio and UV photon flux are relatively low.  

Given that the \ce{CH3OH} ice is likely associated with CO ice 
(since it forms from the latter), then there is also the 
possibility that a fraction of 
\ce{CH3OH} is released during CO ice sublimation at 
$T \gtrsim 17$~K \citep[e.g.,][]{martin-domenech14}.  
In this regime, one would expect that the \ce{CH3OH} emission peaks at the location 
of the CO snowline \citep[$\approx 30$~AU,][]{qi13}, which is suggested 
in these data.  However, this would indicate a different 
chemical origin for gas-phase \ce{CH3OH} compared with \ce{H2O}, 
with the former dependent upon temperature only, and the latter 
dependent upon the UV photon flux (assuming photodesorption as the origin).   
Higher spatial resolution data of multiple methanol transitions 
(to derive excitation information and thus the vertical location) 
will be critical for distinguishing between each chemical origin scenario. 

The detection of gas-phase methanol in a 
protoplanetary disk presents a milestone in the investigation of the molecular 
inventory of disks, and provides much needed constraints on the 
viability of the detection of larger and more complex molecules.  
It also provides an intriguing piece of the puzzle regarding the 
large-scale depletion of \ce{CO} gas in TW~Hya, one of the explanations for which 
is the conversion of CO ice into more complex molecules, such as methanol 
\citep{favre13,kama16,nomura16}.
The detection of cold gas-phase \ce{CH3OH} in TW Hya 
\citep[and warm gas-phase \ce{CH3CN} in MWC~480,][]{oberg15a}
now allows an estimation of  
the likely abundances and resulting line strengths of species 
which are higher up the ladder of complexity.  
Finally, the analysis presented here shows that the high quality 
of the ALMA data allows the stacking of multiple 
transitions to enhance S/N, facilitating the detection and 
analysis of minor gas-phase species in protoplanetary disks.  

\acknowledgments

We thank an anonymous referee for their insightful comments 
which helped improve this paper.  
This work uses the following ALMA data: ADS/JAO.ALMA\#2013.1.00902.S
and  \\ ADS/JAO.ALMA\#2013.1.00114.S. 
ALMA is a partnership of ESO (representing its member states), 
NSF (USA) and NINS (Japan), together with NRC (Canada) and NSC and 
ASIAA (Taiwan), in cooperation with the Republic of Chile. 
The Joint ALMA Observatory is operated by ESO, AUI/NRAO and NAOJ.  
C.~W.~acknowledges support from the 
Netherlands Organization for Scientific Research (NWO, program 639.041.335).  
E.~H.~acknowledges the support of the National Science Foundation for his program in astrochemistry, 
and support from the NASA Exobiology and Evolutionary Biology Program through a subcontract 
from Rensselaer Polytechnic Institute.
Astrophysics at QUB is supported by a grant from the STFC.  
Y.A. acknowledges support from JSPS KAKENHI, grant number 23540266. 
K.I.\"{O}. also acknowledges funding from the Simons Foundation: 
Simons Collaboration on the Origins of Life (SCOL) Investigator award \#321183. 
R.A.L. gratefully acknowledges funding from a National Science Foundation Graduate 
Research Fellowship.
This work is supported by a Huygens fellowship from Leiden University
and by the European Union A-ERC grant 291141 CHEMPLAN.


\end{document}